\documentclass[prl,twocolumn,superscriptaddress,showpacs]{revtex4} 

\usepackage{amsmath}
\usepackage{amsfonts}
\usepackage{verbatim}
\usepackage{subfigure}
\usepackage{sidecap}
\usepackage{color}
\usepackage{epsf,graphicx,amssymb}

\newcommand{\be}{\begin{equation}}
\newcommand{\ee}{\end{equation}}

\begin{document}

\title{Coevolution and correlated multiplexity in multiplex networks}

\author{Jung Yeol Kim}
\affiliation{Department of Physics, Korea University, Seoul 136-713, Korea}
\author{K.-I. Goh}
\email{kgoh@korea.ac.kr}
\affiliation{Department of Physics, Korea University, Seoul 136-713, Korea}
\date{\today}
\begin{abstract}
Distinct channels of interaction in a complex networked system define  network layers, which co-exist and co-operate for the system's function. 
Towards understanding such multiplex systems, we propose a modeling framework based on coevolution of network layers, with a class of minimalistic growing network models as working examples. We examine how the entangled growth of coevolving layers can shape the network structure and show analytically and numerically that the coevolution can induce strong degree correlations across layers, as well as modulate degree distributions. 
We further show that such a coevolution-induced correlated multiplexity
can alter the system's response to dynamical process, exemplified by
the suppressed susceptibility to a social cascade process. 
\end{abstract}

\maketitle

{\em Introduction}---
Agents in complex systems interact in many ways: People are influenced by multiple channels of social interaction such as friendship and work-partnership, and multiple means of transportation such as avian and ground transportations constitute the global transportation infrastructure \cite{szell,multiscale,layered}. Such systems can best be represented as multiplex networks with multiple types of links.  
Each link-type in the system defines a network layer, which is by no means
in isolation but co-exist and co-operate with other layers to fulfill the system's function. Such coupling and interplay of network layers can result in emergent structural and dynamical impact in nontrivial ways \cite{leicht,buldyrev,lee,brummitt,moreno,morris,diffusion}, rendering 
the understanding based on single-network approach incomplete.

In real-world complex systems, either self-organized or man-made, 
the coupling between network layers is not completely random but structured, a property referred to as { correlated multiplexity} \cite{lee}. Specifically, degree of a node (degree is the number of links
a node has) in one layer and that in the other are often strongly correlated. 
Expectedly, she who has many friends tends to have many friendly coworkers in workplace; a hub-airport city is most likely a rail-hub, and so on. Such a structured coupling of network layers
is shown to affect the system's connectivity and robustness properties  \cite{lee,parshani,buldyrev2,NoN}. However, its underlying evolutionary mechanism has not yet been systematically investigated.

In this paper we propose the {\em coevolution} of network layers as an evolutionary mechanism for the correlated multiplexity in growing multiplex networks. To motivate the idea, let us turn back to the transportation network example.  Suppose one were to establish a new air route. In doing so, it might be reasonable to consider not only the candidate city's avian connectivity, but also its ground connectivity such as rail and highway infrastructure in order to maximize the synergy. That is, layers in a multiplex system do not merely co-exist, but they co-evolve, affecting and entangling each other's growth. Elucidating the role of coevolution as a modeling framework of multiplex networks is the main aim of this paper.

To substantiate the key idea, 
we introduce and study a class of minimalistic growing multiplex network models
with coevolving layers based on preferential attachment as working examples. 
We show by analytic calculations assisted
by extensive simulations that coevolution can profoundly affect 
the structure of multiplex systems. 
Not only can it shape the correlated multiplexity, it can also modulate the degree distributions. 
We further demonstrate that multiplex structures with different strength of coevolution respond 
differently to a cascade process, exemplifying the dynamical signature that coevolution can imprint.
Note that coevolution of (single) network structure and dynamical process on it has been studied \cite{holme}. Yet,
coevolution effect of different layers within a multiplex system has remained unexplored.

{\em Modeling framework}---
To enlighten ourselves on the role of coevolution, 
we consider a minimalistic model of coevolving multiplex network  (Fig.~1a).
Each step, a new node enters into the system and in each layer establishes
a link to an existing node. Probability that an existing node would receive
a link from the new node gives the growth kernel $\Pi$ of its degree \cite{ba}.
For degree-based growth, the coevolution of network layers can be formulated
in the way that 
the growth kernel of a node's degree in layer $\mu$ is not only dependent 
on its degree in that layer, $k_{\mu}$, but also on its degrees in other layers  \footnote{One can also formulate coevolution with non-degree based growth, such as static \cite{static} or fitness-based \cite{fitness} growths in a similar manner.}, 
\be
\Pi_{\mu}=f(k_{\alpha},k_{\beta}, \dots, k_{\mu},\dots, k_{\ell}){},\ee
where $\ell$ is the total number of layers in the system 
and we used Greek subscripts to denote the layer index.
We are interested in not only the degree distributions of layers grown under Eq.~(1), 
but also the correlation of degrees across layers to address the correlated multiplexity in the multiplex system. For the latter, 
we calculate Pearson correlation coefficient $\rho_{\mu\nu}$ between the degrees of a node in the two layers $\mu$ and $\nu$ \cite{szell}, given as
\be
{\rho}_{\mu\nu}\equiv\frac{\langle (k-\langle k\rangle)(l-\langle l\rangle)\rangle}{\sigma_k\sigma_l}=
\frac{\langle kl\rangle - \langle k\rangle\langle l\rangle}{\sigma_k\sigma_l}{},
\ee
where we used a simpler notation that $k=k_{\mu}$ and $l=k_{\nu}$ and $\langle k\rangle$ ($\sigma_k$) is the mean (standard deviation)
of node degrees in the given layer.

Henceforth, we focus our analyses on a class of growth kernels based on linear preferential attachment \cite{ba,Dorogovtsev} and systems with $\ell=2$ layers (duplex system) for simplicity, although the main messages would be applicable to more general cases.

\begin{figure}
\centering
\includegraphics[width=.95\linewidth]{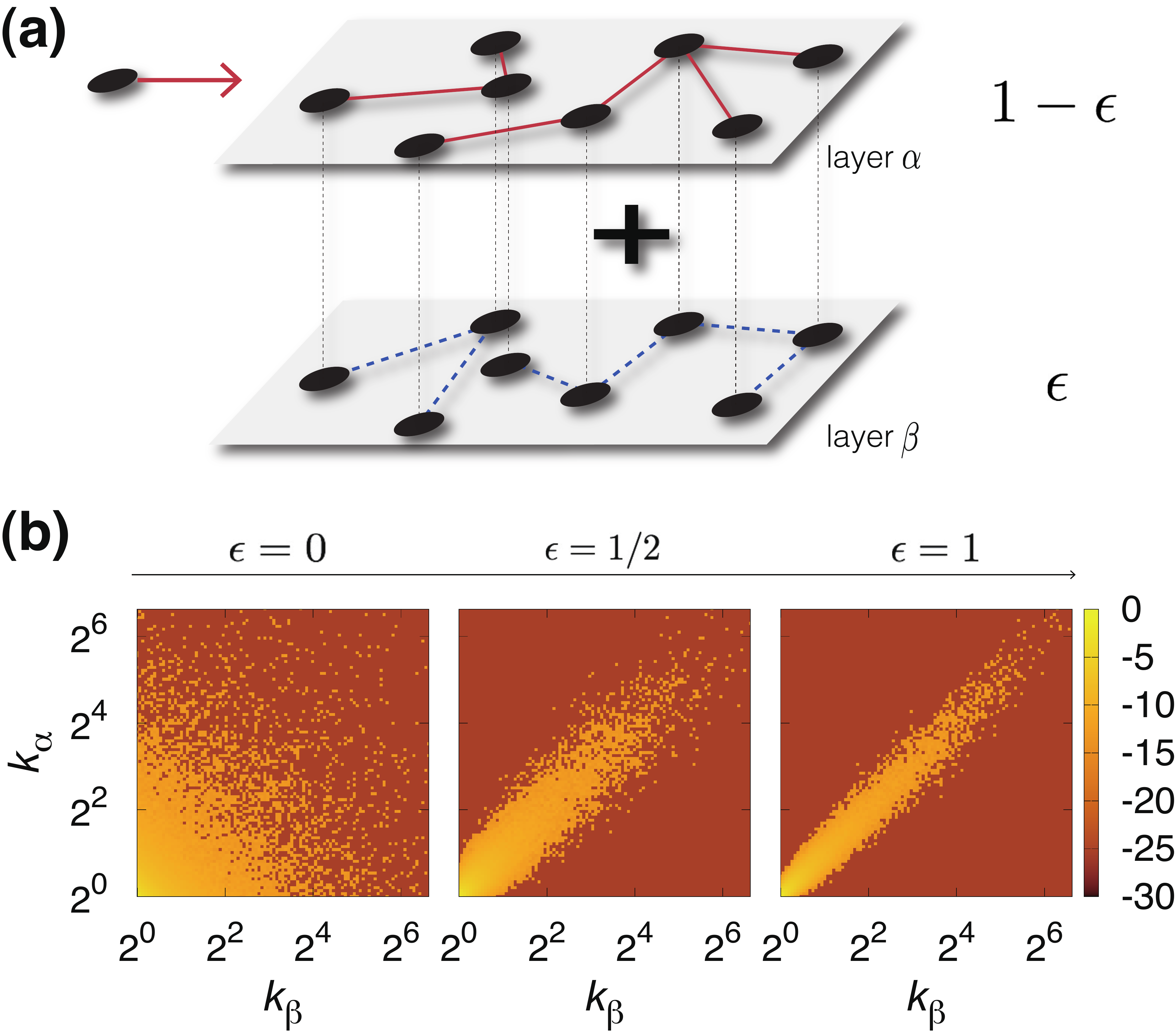}
\caption{(a) Model illustration. Each step, a new node enters the system
and establishes a link in each layer. To choose 
the node to connect in the layer $\alpha$, 
the new node refers to the network connectivity not only in 
that layer $\alpha$ but also in the other layer $\beta$
(and similarly in  the layer $\beta$).
Relative dependency to the other layer is controlled by 
the coevolution factor $\epsilon$.
(b) Numerical simulation results for the joint degree distribution 
$P(k_{\alpha},k_{\beta})$ of coevolving networks of size $N=10^3$
with simple preferential attachment for various $\epsilon$. 
As the coevolution factor $\epsilon$ increases, degrees of a node in the two layers 
become more strongly correlated, intensifying correlated multiplexity. Colorbar denotes
the scale in $\ln P(k_{\alpha},k_{\beta})$.}
\end{figure}

{\em Mutually-dependent layers}--- 
Let us suppose the growth kernels  for the two layers given by 
\begin{subequations}
\begin{align}
\Pi_{\alpha} &\propto [ (1-\epsilon )(k_{\alpha}+a)+\epsilon (k_{\beta}+a )]{}, \\
\Pi_{\beta} &\propto [ \epsilon(k_{\alpha}+a)+(1-\epsilon )(k_{\beta}+a )]{}.
\end{align}
\end{subequations}
Here $\epsilon$ is a parameter that controls the strength of coevolution, hence called the coevolution factor. As $\epsilon>0$ increases, the two layers coevolve with mutually depending more strongly.
$a$ is the shift factor introduced to control the layer's native degree exponent.
Recall that growing network with $\Pi(k) \propto (k+a)$ has an asymptotic power-law degree distribution, $P(k)\sim k^{-\gamma}$, 
with the exponent $\gamma=3+a$~\cite{Dorogovtsev}. 

The case with $a=0$ (simple preferential attachment~\cite{ba}) is particularly illustrative as it is amenable to most detailed analytic results as well as efficient numerical simulations.
The rate equation for node $i$'s degree in layer $\alpha$ take a simple form as
\be
\frac{\;\partial k_{i,\alpha}}{\partial t}= \frac{(1-\epsilon)k_{i,\alpha}+\epsilon k_{i,\beta}}{2t}{},
\ee
and similarly for the layer $\beta$.
The solution of Eq.~(4) takes the same form as the original Barab\'asi-Albert model 
as $k_{i}(t)= \left(t/t_{i}\right)^{1/2}$ for both layers, where $t_i$ is the arrival time of node $i$ \cite{ba}. 
This leads to scale-free network layers with
$P(k) \sim k^{-3}$ for both layers, irrespective of the coevolution factor $\epsilon$. 
The degree correlation is, however, crucially affected by the coevolution
(Fig.~1b).

To see this, we set up 
a rate equation for the number of nodes having degree $k$ on the layer $\alpha$ and $l$ on the layer $\beta$ at time $t$, denoted as $C_{k,l}(t)$,  which reads 
\begin{align}
\frac{dC_{k,l}}{dt}=&\left[\Pi^{(\alpha)}_{k-1,l}\left(1-\Pi^{(\beta)}_{k-1,l}\right)\right] C_{k-1,l} \nonumber\\
&+\left[\Pi^{(\beta)}_{k,l-1} \left(1-\Pi^{(\alpha)}_{k,l-1}\right)\right] C_{k,l-1} \nonumber\\
&+ \left[\Pi^{(\alpha)}_{k-1,l-1} \Pi^{(\beta)}_{k-1,l-1}\right] C_{k-1,l-1} \nonumber\\
&- \left[1-\left(1- \Pi^{(\alpha)}_{k,l}\right)\left(1- \Pi^{(\beta)}_{k,l}\right) \right] C_{k,l} + \delta_{k1}\delta_{l1}{}, 
\end{align}
where the parenthesized superscript is used to denote the layer index.
Changing variable by $C_{k,l}(t) = t c_{k,l}(t)$ and introducing the generating function for $c_{k,l}(t)$ \cite{krapivsky}, 
one can obtain following coupled differential equations for $\langle kl\rangle$ and $\langle k^2\rangle$.
\begin{subequations}
\begin{align}
\left[\epsilon-\frac{\epsilon^2+(1-\epsilon)^2}{4t} \right] & \langle kl \rangle+t\frac{\partial \langle kl \rangle}{\partial t} 
=\epsilon\langle k^2 \rangle+\frac{\epsilon(1-\epsilon)}{2t}\langle k^2 \rangle+1{}, \\
\epsilon\langle kl \rangle &=\epsilon\langle k^2 \rangle+t\frac{\partial \langle k^2 \rangle}{\partial t}-2{}. 
\end{align}
\end{subequations}
Solving for $\langle k^2\rangle$, one obtains 
\be
\langle k^2 \rangle = \left\{ \begin{array}{ll}
2{\ln}(t)-c_1{Ei}\left(-\frac{1}{4t}\right)+c_2  & \quad (0<\epsilon \leq1),\\
2{\ln}(t)+d_1 & \quad (\epsilon=0),
\end{array} \right.
\ee
where $Ei(x)$ is the exponential integral \cite{afken} and $c_1, c_2$ and $d_1$ are constants determined by boundary conditions.

\begin{figure}
\centering
\includegraphics[width=.95\linewidth]{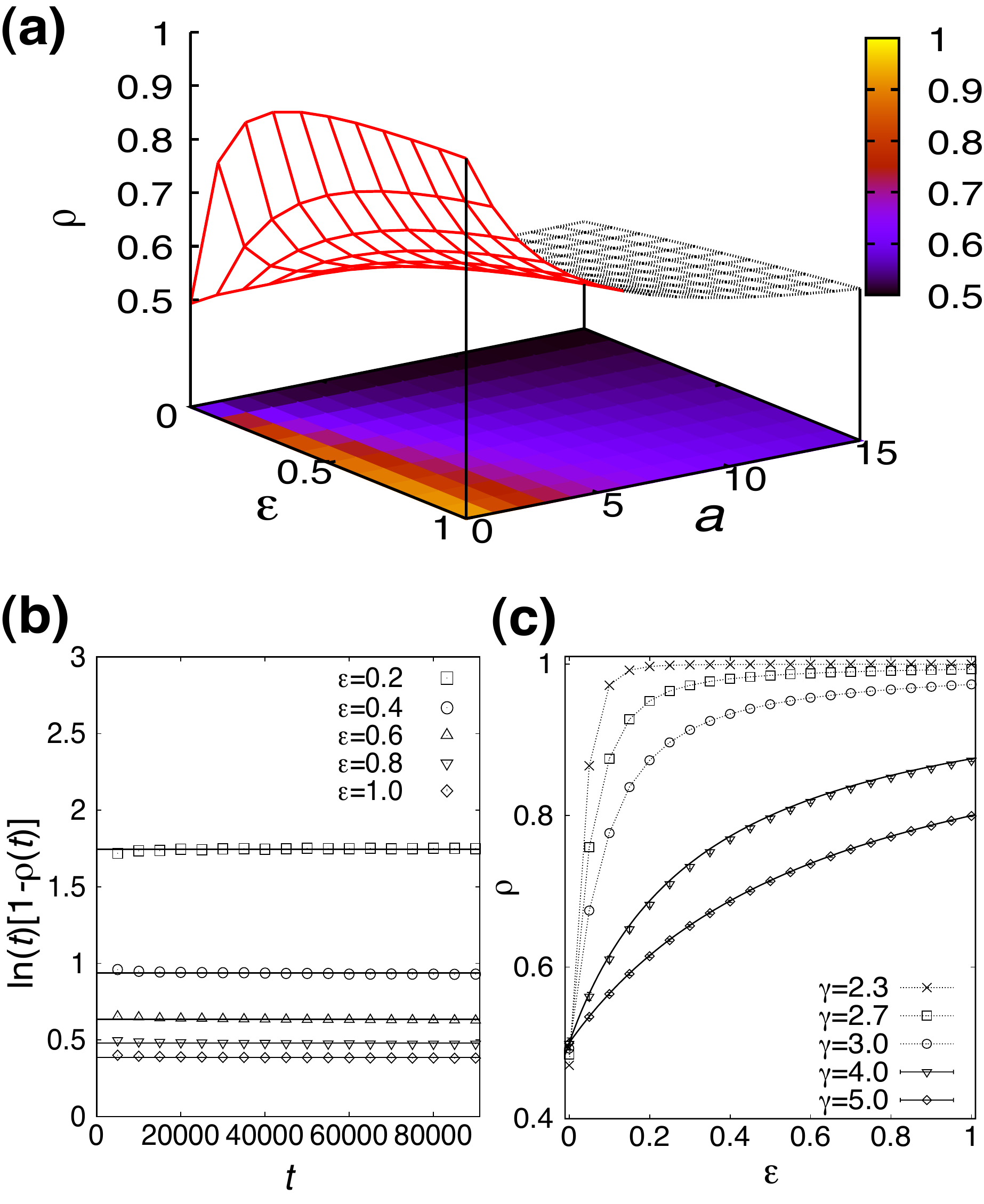}
\caption{Coevolution induces strong correlated multiplexity. (a) 
Degree correlation between two layers $\rho$ as a function 
of coevolutin factor $\epsilon$ and shift factor $a$.
(b) Plots of $\ln(t)[1-\rho(t)]$ as function of $t$ with $a=0$ 
for different $\epsilon$. 
Horizontal behavior of the simulation results (points) supports the analytically-predicted logarithmic convergence of $\rho$ towards the asymptotic value $1$.
The height of horizontal lines gives the coefficient $c$ of logarithmic correction term.
(c) Plots of $\rho$ as a function of $\epsilon$ for various $\gamma$,
obtained from numerical simulations (points) as well as theoretical results Eq.~(10) (solid lines).
Numerical simulations are performed with network size $N=10^6$, averaged over $10^3$ runs.
}
\end{figure}

\begin{subequations}
When $\epsilon=0$, Eqs.~(6) decouple and from Eq.~(6a) one obtains 
\be
\langle kl \rangle = \exp\left(-\frac{1}{4t}\right) \left[c_3-Ei\left(\frac{1}{4t}\right)\right]{},\ee
where $c_3$ is another constant determined by the boundary condition.
For $\epsilon>0$, by plugging Eq.~(7) into Eq.~(6b) we have
\be
\langle kl \rangle=2{\ln}(t)-c_1\left[{Ei}\left(-\frac{1}{4t}\right)-\frac{1}{\epsilon}\exp\left(-\frac{1}{4t}\right)\right]+c_2{}. 
\ee
\end{subequations}
Combining Eqs.~(7) and (8), one can obtain from Eq.~(2) 
the correlation coefficient $\rho$ between degrees of a node in the two layers. 
In the long time (equivalently, large network) limit, the asymptotic value of $\rho$ is obtained as
\be
\rho\to \left\{ 
\begin{array}{ll}
1/2 & \quad (\epsilon=0){}, \\
1 & \quad (0<\epsilon \le 1){}.
\end{array} \right.
\ee

Nonzero correlation even for $\epsilon=0$ can be attributed to the age effect  \cite{parshani} inherent in the growing network, as older nodes have more chance to receive links than newer ones. In that sense,
the network evolution can still be considered coupled even for $\epsilon=0$ as long as the ordering of arrivals of nodes in different layers are correlated as in the present model. A layer's growth becomes completely decoupled and the correlation vanishes ($\rho=0$) only when the arrival times of the same node in different layers are made independent. With coevolution $(\epsilon>0)$, the asymptotic correlation $\rho$ jumps to unity with logarithmically slow convergence, 
as confirmed by numerical simulations~(Fig.~2).

When $a>0$ (the shifted linear kernel),  similar procedure leads to 
$P(k)\sim k^{-(3+a)}$ for both layers. Yet again, the coevolution factor can
affect the correlation.
As $\gamma>3$, $\langle k^2\rangle$ and $\langle kl\rangle$ remain finite as network grows and converge rapidly to the limiting value. It is thus sufficient to focus on the limiting values. Similar but slightly more involved calculations lead to
\be
\rho \to \frac{6\epsilon+a}{6\epsilon+2a}{}, \ee
in excellent agreement with numerical simulations (Fig.~2c). $\rho$ increases with $\epsilon$.
It decreases with $a$ (or equivalently, $\gamma$), yet remains $\rho>1/2$ as long as $\gamma$ is finite (Fig.~2a). 
Similar results are observed for $-1<a<0$, corresponding to $2<\gamma<3$, found in many real-world examples (Fig.~2c).
These results clearly highlight the role of coevolution factor in shaping correlated multiplexity.

{\em Unidirectional dependency and dissimilar kernel}---
Layers may influence each other asymmetrically and non-reciprocally. Furthermore, each layer may have different native growth dynamics.
The former can be dictated by distinct $\epsilon$ parameters for the two layers, and the latter by different $a$ parameters, improving the limitation of monoparametric coupling in Eq.~(3) towards more realistic modeling.
To illustrate the effects of such factors, we consider the growth kernel
of the following form 
\begin{subequations}
\begin{align}
\Pi_{\alpha} &\propto (k_{ \alpha}+a){}, \\
\Pi_{\beta} &\propto \left[\epsilon k_{\alpha}+(1-\epsilon)k_{\beta}\right]{}.\end{align}
\end{subequations}
That is, the layer $\alpha$ grows autonomously with shift factor $a$, 
but the layer $\beta$ evolves with coevolution factor $\epsilon$, 
representing cases with unidirectional dependency with dissimilar growth kernel.

In this case even the degree equation becomes quite involved for layer $\beta$, but the limiting behavior can be obtained that 
\be
P_{\beta}(k) \sim \left\{ \begin{array}{ll}
k^{-{(3-\epsilon)}/{(1-\epsilon)}}&\quad\textrm{($\epsilon \rightarrow 0){},$}\\
k^{-(3+a)}&\quad\textrm{($\epsilon \rightarrow 1$),}
\end{array} \right.
\ee
with the two regimes separated by $\epsilon \simeq a/(2+a)$. 
For the independently evolving layer $\alpha$, $P_{\alpha}(k)  \sim k^{-(3+a)}$. 
This result shows that the degree distribution of the dependent layer $\beta$
becomes modulated by the degree distribution of the layer $\alpha$ it depends on,
if the coevolution factor is strong enough. 
This intriguing analytical prediction is supported by numerical simulations (Fig.~3). 
Asymptotic value of $\rho$ 
for coevolving case ($\epsilon>0$) is obtained as 
\be
\rho\rightarrow \sqrt{\frac{2+a}{2+a+a/\epsilon}}{},
\ee
with the correlation $\rho$ increasing with the coevolution factor $\epsilon$.  
When $\epsilon\to 0$, $\rho$ vanishes asymptotically for any $a>0$, 
which may suggest the asymmetric coupling as a possible factor driving low correlation between
independently evolving layers.

\begin{figure}
\centering
\includegraphics[width=.8\linewidth]{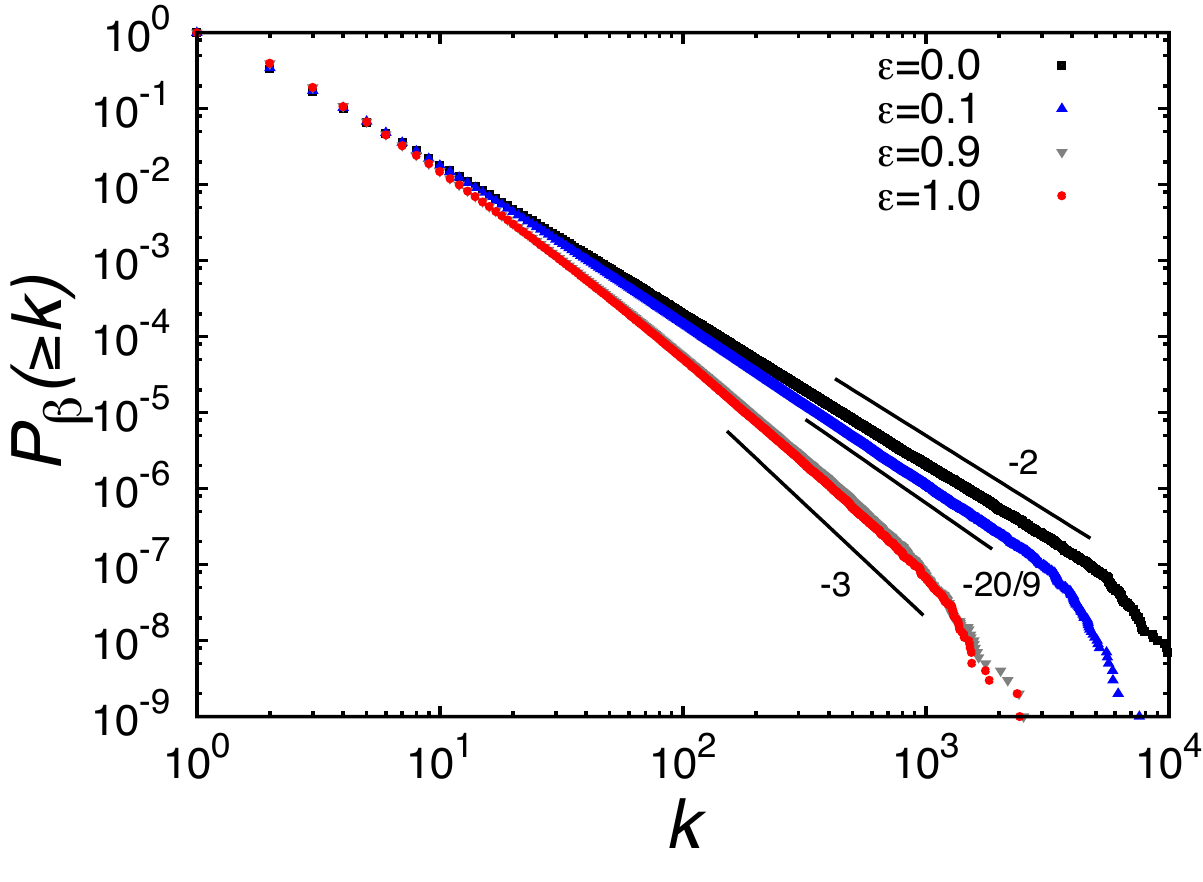}
\caption{
Plotted are the cumulative degree distribution $P_{\beta}(\ge k)$ of the 
dependently-evolving layer $\beta$ for different coevolution factor $\epsilon=0, 0.1, 0.9, 1.0$ (top to bottom) with fixed $a=1$. 
Straight lines indicate theoretical exponents from Eq.~(12), with slopes
$-2, -20/9$, and $-3$ (top to bottom).
Data are obtained from the networks with size $N=10^7$, averaged over $10^2$ runs.
Note the results for $\epsilon=0.9$ and $1.0$ almost overlap.
}
\end{figure}

{\em Impact on cascade dynamics.}--- Finally, we study the effect of 
coevolved multiplex  structure on dynamical processes occurring on it. 
As a specific example, we consider the social cascade model which was introduced by Watts \cite{watts} and recently generalized for multiplex social networks \cite{brummitt}, as the multiplex social network is one of the most actively studied multiplex systems \cite{szell,leicht,brummitt,diffusion}. 
In this model, each node (individual) can be in either active or inactive state. In the original single network version \cite{watts}, an inactive node switches to active state if the fraction of active neighbors exceeds the prescribed threshold $R$. The final fraction $\phi$ of the active nodes in the network, starting from a small fraction $\phi_0$ of initial active seed nodes, measures how susceptible a network is to the cascade process. 
In the multiplex version \cite{brummitt}, a node gets activated if the fraction of active neighbors exceeds the threshold in {\em any} layer, facilitating global cascades to the extent that layers unsusceptible to global cascades in simplex can cooperatively achieve them when multiplex-coupled.
Here we take fraction $\phi_0$ of highest degree nodes as initial active seeds, and measure what fraction $\phi$ of nodes are activated at the end of the multiplex cascade process. 

To highlight the effect of coevolution factor, we first compare the cascade processes on two network structures with $\epsilon=0$ and $\epsilon=1$, respectively. Results for networks with layers growing with linear kernels, Eq.~(4), are shown in Fig.~4.  
For a wide range of threshold $R$, the networks with $\epsilon=1$ support significantly smaller cascades than those with $\epsilon=0$.
For given $R$, the cascade size monotonically decreases with the coevolution factor $\epsilon$ (Fig.~4, inset), showing that the coevolved structure with strong correlated multiplexity can be significantly less susceptible to cascades. 
Note that the superposed network structures in Fig.~4 are independent of $\epsilon$, thus the coevolution factor modulates only internal rearrangement of layer structure. The fact that structural modulation within such a limited range could lead to an observable macroscopic difference in dynamics elucidates the nontrivial role of coevolution.

\begin{figure}
\centering
\includegraphics[width=0.8\linewidth]{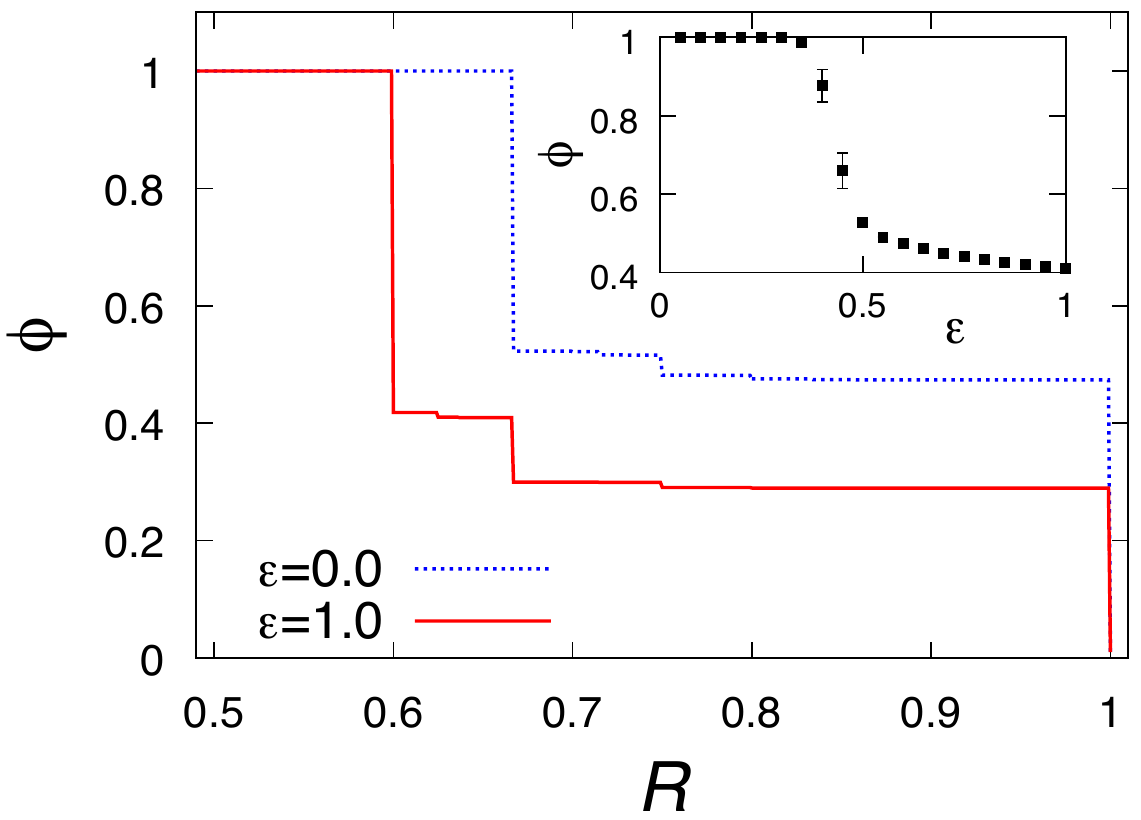}
\caption{Coevolution can suppress the network's susceptibility to cascade process. 
(Main) Compared are the final cascade size $\phi$ as a function of the threshold $R$ with $\phi_0=10^{-2}$,
on networks of size $N=10^5$ with $\epsilon=1$ (red solid) and $\epsilon=0$ (blue dotted).
(Inset) $\phi$ as a function of $\epsilon$ for fixed $R=0.65$,
showing that $\phi$ monotonically decreases with $\epsilon$.}
\end{figure}

{\em Summary}--- To summarize, we have proposed a multiplex network modeling framework based on coevolution of network layers. We have shown both analytically and numerically that the coevolution can profoundly alter the structural properties of the evolved network, both in the degree distribution within the layer and in the degree correlation across the layers. Coevolved multiplex structures spontaneously develop strong correlated multiplexity. Such a structural modulation of coevolved multiplex is further shown to entail dynamical signature, exemplified by the suppressed susceptibility to a cascade process. 

As coevolution of network layers takes place ubiquitously
from social \cite{szell} and infrastructural \cite{buldyrev} 
to economic and ecological systems \cite{eco},
the proposed coevolution-based modeling framework 
could serve as a starting point for further investigation in diverse fields with richer system-specific contexts and details, a rationale shared by a recent independent work by Nicosia {\it et al.}~\cite{nicosia} whose results partly overlap with ours. Several more realistic features such as  difference in number of nodes
or delayed arrivals of a node \cite{nicosia} in different layers would also affect correlation property of multiplex structure. 
Another factor of interest is  
the effect of negative coupling between layers.
These details can be readily incorporated into model variants based on the proposed framework,
which we plan to explore in a follow-up study.

\begin{acknowledgments}
We thank the anonymous Referees for numerous critical and insightful comments.
This work was supported by Basic Science Research Program through the NRF grant funded by MEST (No. 2011-0014191). 
\end{acknowledgments}


\begin{thebibliography}{99}

\bibitem{szell} M. Szell, R. Lambiotte, and S. Thurner, Proc. Natl. Acad. Sci. U.S.A. {\bf 107}, 13636 (2010); M. Szell and S. Thurner, Soc. Netw. {\bf 32}, 313 (2010).
\bibitem{multiscale} D. Balcan {\it et al.}, Proc. Natl. Acad. Sci. U.S.A. {\bf 106}, 21484 (2009).
\bibitem{layered} M. Kurant and P. Thiran, Phys. Rev. Lett. {\bf 96}, 138701 (2006).
\bibitem{leicht} E. A. Leicht and R. M. D'Souza, arXiv:0907.0894.
\bibitem{lee} K.-M. Lee, {\it et al.}, New J. Phys. {\bf 14}, 033027 (2012).
\bibitem{buldyrev} S. V. Buldyrev, {\it et al.}, Nature (London) (2010).
\bibitem{brummitt} C. D. Brummitt, K.-M. Lee, and K.-I. Goh, Phys. Rev. E {\bf 85}, 045102(R) (2012).
\bibitem{moreno} E. Cozzo, A. Arenas, and Y. Moreno, Phys. Rev. E {\bf 86}, 036115 (2012).
\bibitem{morris} R. G. Morris and M. Barthelemy, Phys. Rev. Lett. {\bf 109}, 128703 (2012).
\bibitem{diffusion} S. G\'omez, A. D\'iaz-Guilera, J. G\'omez-Garde\~nes, C. J. P\'erez-Vicente, Y. Moreno, and A. Arenas, Phys. Rev. Lett. {\bf 110}, 028701 (2013).
\bibitem{parshani} R. Parshani, {\it et al.}, EPL {\bf 92}, 68002 (2010).
\bibitem{buldyrev2} S. V. Buldyrev, N. Shere, and G. A. Cwilich, Phys. Rev. E {\bf 83}, 016112 (2011).
\bibitem{NoN} K.-M. Lee, J. Y. Kim, S. Lee, and K.-I. Goh, in {\it Network of networks.}  (eds.) G. D'Agostino and A. Scala (Springer, Heidelberg, to appear).
\bibitem{holme} P. Holme and M. E. J. Newman, Phys. Rev. E {\bf 74}, 056108 (2006); S.-W. Kim and J. D. Noh, Phys. Rev. Lett. {\bf 100}, 118702 (2008).
\bibitem{ba} A.-L. Barab\'asi and R. Albert, Science {\bf 286}, 509 (1999).
\bibitem{Dorogovtsev} S. N. Dorogovtsev, J. F. F. Mendes, and A. N. Samukhin, Phys. Rev. Lett. {\bf 85}, 4633 (2000).
\bibitem{krapivsky} P. L. Krapivsky and S. Redner, Phys. Rev. E {\bf 63}, 066123 (2001).
\bibitem{afken} G. B. Arfken and H. J. Weber, {\it Mathematical Methods for Physicists, 6th ed.} (Academic Press, New York, 2005).
\bibitem{watts} D.~J.~Watts, Proc. Natl. Acad. Sci. U.S.A. {\bf 99}, 5766 (2002).
\bibitem{eco} C. A. Hidalgo and R. Hausmann, Proc. Natl. Acad. Sci. U.S.A. {\bf 26}, 10570 (2009); M. J. O. Pocock, D. M. Evans, and J. Memmott, Science {\bf 335}, 973 (2012).
\bibitem{nicosia} V. Nicosia, {\it et al.}, arXiv:1302.7126.
\bibitem{static} K.-I. Goh, B. Kahng, and D. Kim, Phys. Rev. Lett. {\bf 87}, 278701 (2001).
\bibitem{fitness} G. Caldarelli, A. Capocci, P. De Los Rios, and M. A. Mu\~noz, Phys. Rev. Lett. {\bf 89}, 258702 (2002).

\end{thebibliography}
\end{document}